\documentclass{SciPost}

\hypersetup{
    colorlinks,
    linkcolor={red!50!black},
    citecolor={blue!50!black},
    urlcolor={blue!80!black}
}

\usepackage[bitstream-charter]{mathdesign}
\DeclareSymbolFont{usualmathcal}{OMS}{cmsy}{m}{n}
\DeclareSymbolFontAlphabet{\mathcal}{usualmathcal}

\fancypagestyle{SPstyle}{
\fancyhf{}
\lhead{\colorbox{scipostblue}{\bf \color{white} ~SciPost Physics }}
\rhead{{\bf \color{scipostdeepblue} ~Submission }}

\fancyfoot[C]{\textbf{\thepage}}
}

\usepackage{soul} 

\usepackage[T1]{fontenc}
\usepackage[utf8]{inputenc}
\usepackage{hyperref}  
\usepackage{multirow}  
\usepackage{array}
\usepackage{booktabs} 
\usepackage{siunitx}

\usepackage{bbm}       
\usepackage{bm}        
\usepackage{mathtools} 
\usepackage{braket}    

\usepackage{xcolor} 

\begin{document}

\pagestyle{SPstyle}

\begin{center}{\Large \textbf{\color{scipostdeepblue}{
Reduced density matrix approach to one-dimensional ultracold bosonic systems\\
}}}\end{center}

\begin{center}\textbf{
Mitchell J. Knight\textsuperscript{1$\star$},
Harry M. Quiney\textsuperscript{1} and 
Andy M. Martin\textsuperscript{1}
}\end{center}

\begin{center}
{\bf 1} School of Physics, University of Melbourne, Parkville, 3010, Australia
\\[\baselineskip]
$\star$ \href{mailto:knightm1@student.unimelb.edu.au}{\small knightm1@student.unimelb.edu.au}\,,\quad
\end{center}

\section*{\color{scipostdeepblue}{Abstract}}
\textbf{\boldmath{%
	The variational determination of the two-boson reduced density matrix is described for a one-dimensional system of $N$ harmonically trapped bosons interacting via contact interaction, where $N$ ranges from $2$ to $10^4$. The ground-state energies are calculated, and compared to existing methods, including the analytic case (for $N=2)$ and mean-field approaches such as the one-dimensional Gross-Pitaevskii equation and its variations. Structural properties such as the density and correlation functions are also derived, including the behaviour of the correlation function when boson coordinates coincide. This collectively demonstrates the capacity of the reduced density matrix method to accurately calculate collective ground-state properties of bosonic systems comprising few to many bosons, including the cross-over region between these extremes, across a large range of interaction strengths.
}}

\vspace{\baselineskip}

\noindent\textcolor{white!90!black}{%
\fbox{\parbox{0.975\linewidth}{%
\textcolor{white!40!black}{\begin{tabular}{lr}%
  \begin{minipage}{0.6\textwidth}%
    {\small Copyright attribution to authors. \newline
    This work is a submission to SciPost Physics. \newline
    License information to appear upon publication. \newline
    Publication information to appear upon publication.}
  \end{minipage} & \begin{minipage}{0.4\textwidth}
    {\small Received Date \newline Accepted Date \newline Published Date}%
  \end{minipage}
\end{tabular}}
}}
}


\vspace{10pt}
\noindent\rule{\textwidth}{1pt}
\tableofcontents
\noindent\rule{\textwidth}{1pt}
\vspace{10pt}


\section{Introduction}

	Systems of ultracold bosons have long been a topic of theoretical and experimental research. The two-particle interaction in such systems is parametrised by a single parameter known as the $s$-wave scattering length, $a_s$, which is tunable via Feshbach resonances \cite{Feshbach1958}. These systems are highly controllable and serve as excellent test-beds to study the ground-state and dynamics of few- to many-body quantum mechanics. Despite recent experimental success in the study of bosonic gases \cite{Anderson1995,Gorlitz2001,Kinoshita2004}, their theoretical study long pre-dates their experimental realisation. Such systems, in particular the lower-dimensional cases considered here, exhibit regimes that are well-described by integrable models, readily allowing for the derivation of approximate forms of the wavefunction. 
	
	A special case of particular interest is that of a one-dimensional system of bosonic particles interacting via a contact interaction, similarly parametrised by a modified form of the $s$-wave scattering length. Such systems are described by simple Hamiltonians, with the interaction described by a Dirac $\delta$-function. Moreover, such systems are often considered to be harmonically trapped or, in the absence of such a trap, constrained under well-defined boundary conditions, which are typically periodic. These systems were first introduced by Tonks \cite{Tonks1936} and later studied in detail by Girardeau \cite{Girardeau1960} and Lieb and Liniger \cite{Lieb1963}. The Lieb-Liniger (LL) model is well-studied and is one of the simplest Hamiltonians that is, in principle, exactly solvable  via a Bethe \textit{ansatz}. In systems described by the LL model and its counterpart involving a harmonic trapping potential, a complex phenomenology emerges. For example, in the limit that the interactions dominate, a "traffic-jam" phenomenon appears whereby bosons cannot pass each other, leading to a "fermionisation" of the system \cite{Girardeau1960}. Such a physical system is known as a Tonks-Girardeau (TG) gas, with the ground-state described by mapping the $N$-particle bosonic wavefunction to that of a $N$-particle \textit{fermionic} wavefunction for a non-interacting system. 
	
	Generally speaking, the LL model allows for arbitrary interactions that are typically repulsive and is generally described in the absence of any trapping potential. Notably, the TG gas is the limit of the LL model in which the interaction strength becomes arbitrarily large. Another system of particular importance is the one-dimensional Bose-Einstein condensate (1D BEC), which consists of a system of harmonically trapped bosons interacting weakly via the contact interaction. These systems are models of quasi-1D BECs which can be achieved with extreme, prolate (cigar-shaped) trapping geometries. They differ in their characteristics from the extreme example of a TG gas or, generally, a LL model of bosons. Moreover, they are well-described for large numbers of particles by a reduced form of the semi-classical three-dimensional Gross-Pitaevskii equation (3D GPE), known as the one-dimensional Gross-Pitaevskii equation (1D GPE) \cite{Gross1961,Pitaevskii1961,Salasnich2002,Bao2012}.
	
	In the following, we consider a one dimensional system of harmonically trapped bosons interacting via a contact interaction. As indicated above, a range of different theoretical techniques has been developed to study such systems. However, there remains a lack of a single, computationally tractable methodology that accurately spans the full parameter space of particle number (from few to many) and interaction strength (both repulsive and attractive) without significant approximations or regime-specific adjustments. For example, an analytic solution for harmonically trapped bosons only exists for $N=2$ \cite{Busch1998}, where $N$ is the number of bosons. Directly diagonalising the Hamiltonian within a limited basis set can offer highly accurate results for systems with $N>2$, but this method scales pathologically with $N$ and is practical only for systems with $N\lesssim 10$. A poor scaling in computational complexity is not limited to direct diagonalisation, however, as other approaches such as quantum Monte Carlo \cite{Krauth1996,Purwanto2004} and methods arising from quantum chemistry, including the coupled cluster method \cite{Cederbaum2006,Bhowmik2024} all scale polynomially in the size of the system, which depends upon $N$. The Bethe \textit{ansatz} approach, whilst benefiting from linear scaling in $N$, is limited in its application only to integrable systems where the underlying Hamiltonian is translationally invariant; in this work, we consider trapped systems which are not translationally invariant. These approaches are intractable for large systems (where $N>100$) and mean-field approaches are more suitable, including Hartree-Fock theory \cite{Hartree1928a,Hartree1928b,Fock1930,Fock1930a,Esry1997,Bergeman1997,Ohberg1998} and the 1D GPE. These are only valid in the mean-field limit where the correlations in the system are small  and are thus ill-equipped to describe bosonic systems with fewer particles or strong interactions.
	
	These methods all aim to calculate the $N$-particle wavefunction directly, using a variety of technical approaches. However, different approaches using \textit{reduced} quantities, such as the density, or the diagonal component of the one-particle reduced density matrix (1-RDM), are well-described, such as in reduced density matrix functional theory (RDMFT) \cite{Liebert_2021, Liebert_2023}. A significant benefit is the lack of scaling with particle number, yet such a benefit is offset by a lack of knowledge of the exact energy functional; notably, there has been recent success in applying RDMFT to bosonic systems \cite{Benavides-Riveros2020}.
	
	Considering the above, we are justified in placing the common theoretical approaches into two broad categories: size-extensive, which can describe weakly and strongly interacting systems of few bosons, and \textit{not} size-extensive, which can describe weakly interacting systems of many-particles. As such, there remains an apparent absence of a theoretical approach that can describe both broad categories as well as the crossover behaviour from few- to many-body systems. 
		
		A promising approach lies in the application of the two-particle reduced density matrix (2-RDM). As the Hamiltonian itself depends only on pairwise interactions, the energy of the system can be described exactly as the trace of the product of the 2-RDM and the Hamiltonian, allowing one to use the 2-RDM, which is considerably less complex than the wavefunction, as the trial object in a variational scheme to minimise the energy. The 2-RDM method has been applied extensively in quantum chemistry \cite{Mazziotti2002,Mazziotti2012,Zhao2004,Nakata2001,Fukuda2007,Nakata2008,Mazziotti2012c,Mazziotti2004,Mazziotti2004a,Mazziotti2011,Rosina1975,DePrince2010,Maradzike2020,DePrince2008,Mazziotti2020,Head2020}, and to more esoteric systems such as Hubbard model-based systems \cite{Anderson2013,Hammond2006,Verstichel2012} and ultracold, few-fermion systems \cite{Knight2022}. Primarily, the utilisation of the 2-RDM method has been for fermionic systems, yet bosonic systems have also been studied, such as the simple one-dimensional system with harmonic interactions \cite{Gidofalvi2004}. In all cases, the 2-RDM must be constrained by a set of conditions that ensure it is derivable from a legitimate $N$-particle density matrix (whether mixed or pure) and hence a physical quantum state. These are the so-called $N$-representability conditions which have been well-studied, both in fermionic and bosonic contexts \cite{Gidofalvi2004,Coleman1963,GarrodPercus1964,Mazziotti2012a,Mazziotti2012b,Mazziotti2006}. 
		
		As the 2-RDM method scales favourably with particle number, it is enticing to leverage this approach to describe one-dimensional bosonic systems across a large parameter space spanned by $N$, from very few to very many. Accordingly, in this work we apply the 2-RDM methodology to a one-dimensional systems of $N$, harmonically trapped bosons which interact by a simple, Dirac $\delta$-function contact interaction. The ground-state energies, densities, and correlations are calculated, along with demonstrations of the method's universality, being equally applicable to small and large systems, regardless of the strength of the repulsive interactions.

\section{Theory}
	
	\subsection{The Reduced Density Matrix}
	
		Given an arbitrary $N$-boson (or generic particles of arbitrary spin, more generally) system in 1D, 2D, or 3D, with pairwise interactions, the many-body Hamiltonian will take the form,
		\begin{equation}
			\hat{\mathcal{H}}\coloneqq \sum_{i=1}^N\,{}^{(1)}\hat{h}(\bm x_i)+\frac{1}{2}\sum_{\substack{i,j=1 \\ i\neq j}}^N\,{}^{(2)}\hat{\mathcal{V}}(\bm x_i,\bm x_j),\label{equation:many_body_hamiltonian}
		\end{equation}
		where ${}^{(1)}\hat{h}(\bm x_i)$ is a one-particle operator and ${}^{(2)}\hat{\mathcal{V}}(\bm x_i,\bm x_j)$ is a two-particle operator, with $\bm x_i$ being the coordinates for the $i^{\rm th}$ boson. Typically, variational approaches are utilised to find the $N$-particle wavefunction, $\Psi (\bm x_1,\bm x_2,\dots,\bm x_N)$ by means of the Schr\"odinger equation. However, it can be demonstrated \cite{Husimi1940,Lowdin1955,Coleman1963,terHaar1961,McWeeny1960} that the energy of a system of particles interacting pairwise can be expressed as a linear functional of the 1- and 2-RDMs, defined as

			\begin{align}
				{}^{(1)}D(\bm x_1;\bm x_1')					&\coloneqq	N\int\text{d}\bm x_2\text{d}\bm x_3\cdots\text{d}\bm x_N\,\Psi(\bm x_1,\bm x_2,\dots,\bm x_N)\Psi^*(\bm x_1',\bm x_2,\dots,\bm x_N),\\
				{}^{(2)}D(\bm x_1,\bm x_2;\bm x_1',\bm x_2')	&\coloneqq	N(N-1)\int\text{d}\bm x_3\text{d}\bm x_4\cdots\text{d}\bm x_N\,\Psi(\bm x_1,\bm x_2,\bm x_3,\dots,\bm x_N)\Psi^*(\bm x_1',\bm x_2',\bm x_3,\dots,\bm x_N).
			\end{align}
The normalisations are by choice, with the unit normalisation also appearing prominently in the literature. Through the introduction of an orbital (i.e., single-particle) basis, $\{\varphi_j(\bm x_i)\}$, the 1- and 2-RDMs can be expressed directly as the spectral decompositions,
		\begin{align}
			^{(1)}D(\bm x_1;\bm x_1')					&=\sum_{ij}\,^{(1)}D^i_j\,\varphi_i^*(\bm x_1)\,\varphi_j(\bm x_1'),\label{equation:1rdm_decomposition}\\
			^{(2)}D(\bm x_1,\bm x_2;\bm x_1',\bm x_2')	&=\sum_{ijkl}\,^{(2)}D^{ij}_{kl}\varphi_i(\bm x_1)^*\varphi_j(\bm x_2)^*\varphi_k(\bm x_1')\varphi_l(\bm x_2').\label{equation:2rdm_decomposition}
		\end{align}
		In Eqs.~\eqref{equation:1rdm_decomposition} and \eqref{equation:2rdm_decomposition} the tensors $^{(1)}D^i_j$ and $^{(2)}D^{ij}_{kl}$ can be expressed conveniently in second-quantised notation, providing the most common expressions for the 1- and 2-RDMs as found in the literature,
		\begin{align}
			^{(1)}D^i_j			&=\braket{\Psi|\hat{a}^{\dagger}_i\hat{a}_j|\Psi},\label{equation:1rdm_second_quantised}\\
			^{(2)}D^{ij}_{kl}	&=\braket{\Psi|\hat{a}^{\dagger}_i\hat{a}^{\dagger}_j\hat{a}_l\hat{a}_k|\Psi},\label{equation:2rdm_second_quantised}
		\end{align}
		where $\hat{a}^{\dagger}$ and $\hat{a}$ are the creation and annihilation operators for bosons. We also note that we can express the general Hamiltonian in Eq.~\eqref{equation:many_body_hamiltonian} in a second-quantised form as
		\begin{equation}
			\hat{\mathcal{H}}=\sum_{ij}\,^{(1)}h^i_j\hat{a}^{\dagger}_i\hat{a}_j+\frac{1}{2}\sum_{ijkl}\,^{(2)}\mathcal{V}^{ij}_{kl}\,\hat{a}^{\dagger}_i\hat{a}^{\dagger}_j\hat{a}_l\hat{a}_k,\label{equation:many_body_hamiltonian_second_quantised}
		\end{equation}
		where the indices $i,j,k$ and $l$ correspond to elements of the orbital basis set $\{\varphi_i(\bm x)\}$, the sums run over all possible values of the indices in the basis, $i,j,k,l\in\{1,2,\dots,K\}$, and $^{(1)}h^i_j$ and $^{(2)}\mathcal{V}^{ij}_{kl}$ are the `one-boson integrals' and `two-bosons integrals'.  This nomenclature has been adopted from electronic structure theory, where the equivalent integrals are called `one- and two-electron integrals'
		\begin{align}
			^{(1)}h^i_j			&=	\braket{i|\,^{(1)}\hat{h}|j}=\int\text{d}\bm x\,\varphi_i^*(\bm x)\,^{(1)}\hat{h}\,\varphi_j(\bm x),\label{equation:one_fermion_integral}\\
			^{(2)}\mathcal{V}^{ij}_{kl}	&=	\braket{ij|\,^{(2)}\hat{V}|kl}=\int\text{d}\bm x_1\,\text{d}\bm x_2\,\varphi_i^*(\bm x_1)\varphi_j^*(\bm x_2) \,^{(2)}\hat{\mathcal{V}}\,\varphi_k(\bm x_1)\varphi_l(\bm x_2). \label{equation:two_fermion_integral}
		\end{align}
		By taking the expectation value of the Hamiltonian in Eq.~\eqref{equation:many_body_hamiltonian_second_quantised} with respect to the ground-state $\ket{\Psi}$ and incorporating the expressions of the 1- and 2-RDMs in second-quantised notation (Eqs.~\eqref{equation:1rdm_second_quantised} and \eqref{equation:2rdm_second_quantised}) we see that the ground-state energy is a linear functional of the 1- and 2-RDMs;
		\begin{align}
			\mathcal{E}	&\coloneqq \braket{\Psi|\hat{\mathcal{H}}|\Psi}\\
						&=	\sum_{ij}\,^{(1)}h^i_j\,^{(1)}D^i_j+\frac{1}{2}\sum_{ijkl}\,^{(2)}\mathcal{V}^{ij}_{kl}\,^{(2)}D^{ij}_{kl}\notag\\
						&=	\mathsf{Tr}\left\{\,^{(1)}\hat{h}\,^{(1)}\hat{D}\right\}+\frac{1}{2}\mathsf{Tr}\left\{\,^{(2)}\hat{\mathcal{V}}\,^{(2)}\hat{D}\right\}.\label{equation:energy_rdm}
		\end{align}
		We note that the 1- and 2-RDM are related directly by a trace relation  and hence the energy itself can be expressed directly in terms of the 2-RDM and a slightly modified Hamiltonian \cite{Mazziotti2002,Mazziotti2012}, but here we keep the 1- and 2-RDM contributions to the energy separate. Generally, there are additional constraints on the 1- and 2-RDM during a variational calculation, including trace and spin constraints, and these are covered in detail elsewhere \cite{Zhao2004,Nakata2001,Fukuda2007,Nakata2008,Knight2022}. Notably, the spin constraints do not apply in the bosonic case.

		Crucially, however, are the ensemble $N$-representability conditions that must be enforced during a calculation, which ensure (at least approximately) that the trial 2-RDM is derivable from an ensemble of legitimate $N$-boson density matrices, and hence a physical quantum state, whether mixed or pure. Early calculations involving the 2-RDM proved to yield highly inaccurate results, with the ground-state energy being far \textit{below} expected values \cite{Mayer1955,Tredgold1957}. This is due to the fact that it is not a given that a trial variational 2-RDM is derivable from an ensemble of $N$-particle density matrices, and hence from a physical state. In other words, the resultant 1- and 2-RDMs derived from these relatively unconstrained variational schemes were non-physical. Additional constraints are therefore required on the 1- and 2-RDMs during variational calculations, as first identified by Tredgold \cite{Tredgold1957}. These conditions are known as the \textit{$N$-representability} conditions and the search for them was known as the \textit{$N$-representability problem}. This problem and the derivation of the $N$-representability conditions is discussed at length elsewhere \cite{Coleman1963,Mazziotti2012,Mazziotti2012a,Mazziotti2012b,Mazziotti2012c,Knight2022} and, in this work, we will only provide the $N$-representability conditions used in this work.
	
		$N$-representability conditions manifest practically as constraints that enforce linear operators to be positive semidefinite. Perhaps the simplest of these correspond to constraining the 1- and 2-RDMs themselves to be positive semidefinite. This necessarily enforces the positivity of the eigenvalues of these operators which naturally ensures the probability distributions of single bosons and boson pairs, respectively, to be non-negative. These conditions can be expressed as,
		\begin{align}
			\braket{\Psi|\hat{a}_i^{\dagger}\hat{a}_j|\Psi}	=\,{}^{(1)}D^i_j	&, \quad {}^{(1)}\hat{D}\succcurlyeq0\label{equation:nrep_d1},\\
			\braket{\Psi|\hat{a}_i^{\dagger}\hat{a}_j^{\dagger}\hat{a}_l\hat{a}_k|\Psi}=\,^{(2)}D^{ij}_{kl}	&, \quad {}^{(2)}\hat{D}\succcurlyeq0.\label{equation:nrep_d2}
		\end{align}
		where $\succcurlyeq0$ means `is positive semidefinite.' Eq.~\eqref{equation:nrep_d2} is known as the \textsf{D} condition in the literature. We note, with the condition on the 1-RDM in Eq.~\eqref{equation:nrep_d1}, the 1-RDM is completely $N$-representable, with the eigenvalues and hence the occupation numbers for single-boson states constrained to be non-negative. If we were discussing fermions, this condition alone would not be sufficient as it could lead to violations of the Pauli exclusion principle through instances of the occupation numbers exceeding unity. In such cases, the distribution of \textit{holes} must be constrained in precisely the same manner. This complexity for fermions occurs with the 2-RDM also, whereby the distribution of \textit{two} holes must be constrained in the same manner as the distribution of two particles (the latter of which is manifest in Eq.~\eqref{equation:nrep_d2}). Constraining the distribution of two holes is the well-known \textsf{Q} condition \cite{GarrodPercus1964}. However, this is not the case for bosons so the \textsf{Q} condition is not implemented in this work.
	
		A crucial $N$-representability condition on the 2-RDM, known as the \textsf{G} condition, however, does need to be implemented. This constrains the distribution for one hole and one boson to be non-negative, and can be expressed by,
		\begin{align}
			\braket{\Psi|\hat{a}_i^{\dagger}\hat{a}_j\hat{a}_l^{\dagger}\hat{a}_k|\Psi}	&=	\,^{(2)}G^{ij}_{kl},\\
																						&= {}^{(2)}D^{il}_{kj}+\delta^j_l\,{}^{(1)}D^i_k\\
																						{}^{(2)}\hat{G}&\succcurlyeq0\label{equation:nrep_2rdmc},
		\end{align}
		where the decomposition into the 1- and 2-RDMs is made through the application of the bosonic commutation relations. Additionally, the \textsf{T1} and \textsf{T2} conditions, which correspond to constraining the probability distributions of combinations of three particles and holes to be non-negative, are typically considered in fermionic applications of the RDM methodology. These are discussed at length elsewhere \cite {Zhao2004,Nakata2008,Knight2022} and are not applied in this work. It is worth pointing out that in the fermionic cases, the \textsf{T1} and \textsf{T2} conditions are necessary for highly precise descriptions of the ground state for systems with large correlations, yet in the following we demonstrate that only the simpler (and significantly more efficient in implementation) \textsf{D} and \textsf{G} conditions suffice to extract accurate structural representations of the ground-state. Such a property of bosonic systems was also demonstrated by Gidofalvi \cite{Gidofalvi2004}, where the \textsf{D} and \textsf{G} conditions showed excellent agreement with analytic results for large numbers of bosons interacting harmonically.

		The $N$-representability conditions can be expressed as linear constraints on the elements of the 2-RDM through repeated use of the commutation relations for bosonic creation and annihilation operators. Semidefinite programming techniques \cite{Vandenberghe1996,Gartner2012,Fukuda2007} can then be used to minimise the energy, Eq.~\eqref{equation:energy_rdm}, through variation of the elements of the 2-RDM whilst enforcing the $N$-representability conditions above \cite{Rosina1975,Fukuda2007,Nakata2008,Mazziotti2004,Mazziotti2004a,Mazziotti2011,Mazziotti2012c}.  
	
		In addition to the $N$-representability conditions above, we also implement standard restrictions on the 1- and 2-RDMs during the minimisation of the energy, namely the trace and contraction conditions, expressed as,
		\begin{align}
			\sum_{i}\,{}^{(1)}D^i_i			&=	N,\\
			\sum_{ij}\,{}^{(2)}D^{ij}_{ij}	&=	N(N-1),\\
			\sum_j\,{}^{(2)}D^{ij}_{kj}		&=	(N-1)\,{}^{(1)}D^i_k.
		\end{align}
		In this work, we employ the Semidefinite Programming Algorithm (SDPA) \cite{Yamashita2010,Yamashita2012} which utilises interior point methods, as well as SDPNAL$+$ \cite{Zhao2010,Yang2015,Defeng2020} which uses an augmented Lagrangian method \cite{Zhao2010,Sun2008}, rendering it extremely efficient for large-scale SDPs. To implement these programs, we utilised the Spartan high-performance computing system at the University of Melbourne \cite{Meade2017}.

	\subsection{Ultracold Bosonic Systems}
	
		The system under consideration comprises $N$ bosons at zero temperature, confined by an external trapping potential ${}^{(1)}\hat{\mathcal{V}}_{\mathsf{trap}}(x)\coloneqq m\omega^2x^2/2$. We adopt natural units where $\hbar=m=\omega=1$ in the standard manner. The contact interactions are described by a Dirac $\delta$-function, yielding the simple one- and two-body operators
		\begin{align}
			{}^{(1)}	\hat{h}(x_i)	&=-\frac{1}{2}\partial_{x_i}^2+\frac{1}{2}x_i^2,\label{eq:one_body_ultracold}\\
			{}^{(2)}\hat{\mathcal{V}}(x_i,x_j)	&=g_{\mathsf{1D}}\delta(x_i-x_j).	\label{eq:two_body_ultracold}	
		\end{align}		
		The value of $g_{\mathsf{1D}}$ in an experimental context can be determined, and is described in the next section, but it suffices to state here that it is simply a real number that parametrises the `strength' of the interaction. A convenient basis to study this system is the set of one-dimensional quantum harmonic oscillator eigenfunctions which are defined by
		\begin{equation}
			\varphi_n(x)\coloneqq \frac{1}{\pi^{1/4}}\frac{1}{\sqrt{2^nn!}}\exp\left(-\frac{x^2}{2}\right)\,H_n(x),\label{eq:basis}
		\end{equation}
		where $H_n(x)$ is a Hermite polynomial. The one- and two-boson integrals are both analytic (with the two-bosons integrals also being calculated exactly via Gauss-Hermite quadrature). 
	
	\subsection{The Gross-Pitaevskii Equation and its Dimensionally Reduced Form}
	
		The Hamiltonian described by the one- and two-body operators in Eq.~\eqref{eq:one_body_ultracold} and Eq.~\eqref{eq:two_body_ultracold} is the many-body Hamiltonian corresponding to a Bose-Einstein condensate when $g_{\mathsf{1D}}$ is small. As such, it is worth briefly discussing the 1D GPE, which is the semi-classical approach used in deriving an approximate eigenfunction of this Hamiltonian with the assumption that all bosons are in the lowest energy state.
		
		First, we consider a three-dimensional $N$-boson system at zero temperature, confined by an external trapping potential ${}^{(1)}\hat{\mathcal{V}}_{\mathsf{trap}}(\bm r)$. The van der Waals interactions between the constituent bosons are simplified via the utilisation of the Fermi pseudopotential, as represented by the three-dimensional Dirac $\delta$-function. Such an approach is appropriate for dilute BECs at (near) zero temperature, as the de Broglie wavelength of the bosons is far longer than the range of the boson-boson interaction. For this system, the condensed state, $\psi(\bm r,t)$, satisfying $\int\text{d}\bm r\,|\psi(\bm r,t)|^2=1$, is described by the three-dimensional GPE (3D GPE) \cite{Gross1961,Pitaevskii1961}
		\begin{equation}
			i\hbar\partial_t\psi(\bm r,t)=\left\{-\frac{\hbar^2}{2m}\triangle+{}^{(1)}\hat{\mathcal{V}}_{\mathsf{trap}}(\bm r)+g_{\mathsf{3D}}(N-1)|\psi(\bm r,t)|^2\right\}\psi(\bm r,t),\label{equation:3D_GPE}
		\end{equation}
		where $m$ is the boson mass, $\triangle$ is the Laplacian, $g_{\mathsf{3D}}\coloneqq 4\pi\hbar^2a_{\mathsf{s}}/m$ is a measure of the `strength' of the contact interaction, and $a_{\mathsf{s}}$ is the $s$-wave scattering length.
		
		The corresponding expression in one dimension can be derived by assuming a trapping potential of the form ${}^{(1)}\hat{\mathcal{V}}_{\mathsf{trap}}(\bm r)=m\omega_{\perp}^2\,(x^2+y^2)/2+m\omega_z^2z^2/2$ where $\omega_{\perp}>\omega_z$. In such a case, the wavefunction for the transverse (i.e., $x$- and $y$-directions) can be assumed to exhibit a normalised Gaussian form giving us $\psi(\bm r,t)\sim \exp(-(x^2+y^2)/(2a_{\perp}^2))\,\varphi(z,t)$, where $\varphi(z,t)$ is the wavefunction component defined along $z$ \cite{Salasnich2002}. Then, the 1D GPE can be derived by integrating out the $x$- and $y$-components of the action functional from which the 3D GPE is derived (see Salasnich \textit{et al} \cite{Salasnich2002} for a derivation). In such a case, our wavefunction is assumed to be weakly interacting, and in this limit, we obtain the 1D GPE,
		\begin{equation}
		i\hbar\partial_t\varphi=\left\{-\frac{\hbar^2}{2m}\triangle_z+\frac{1}{2}m\omega_z^2z^2+\frac{g_{\mathsf{3D}}(N-1)}{2\pi a_{\perp}^2}|\varphi|^2\right\}\varphi,
		\end{equation}
		where $a_{\perp}^2\coloneqq \hbar/(m\omega_{\perp})$ is the width of the condensate in the transverse direction and where we note that we remove the explicit appearance of the $(z,t)$ arguments of $\varphi$ for simplicity.
	
	The 1D GPE above exhibits significant accuracy in describing 1D BECs in the weakly interacting limit. However, there does exist more accurate expressions, such as the 1D non-polynomial Schr\"odinger equation (1D NPSE) derived by Salasnich \textit{et al} \cite{Salasnich2002}. The 1D NPSE is still a mean-field description of the state, yet it assumes a variational width in the transverse direction and, accordingly, is more accurate than the 1D GPE. The explicit expression for the 1D NPSE is given by \cite{Salasnich2002},
	\begin{align}
		i\hbar\partial_t\varphi &=	\left\{-\frac{\hbar^2}{2m}\triangle_z+\frac{1}{2}m\omega_z^2z^2+\frac{\hbar^2}{2m\sigma^2}+\frac{m\omega_{\perp}^2\sigma^2}{2}+\frac{g_{\mathsf{3D}}(N-1)}{2\pi \sigma^2}|\varphi|^2\right\}\varphi,\label{equation:npse_differential} \\
						0			&=\frac{\hbar^2}{2m\sigma^3}-\frac{1}{2}m\omega_{\perp}^2\sigma+\frac{g_{\mathsf{3D}}(N-1)|\varphi|^2}{4\pi\sigma^3}.\label{equation:npse_algebraic}
	\end{align}
	Here, $\sigma$ is the variational width of the BEC in the transverse direction. We note that in Eq.~\eqref{equation:npse_algebraic}, we can solve for $\sigma$ and then substitute the relevant expression into the differential equation in Eq.~\eqref{equation:npse_differential}. Whilst the 1D NPSE is formally derived from the 3D GPE by assuming a  transverse profile with a variational width $\sigma(z)$, its application to strictly 1D systems is well justified in the present context. The ground-state solution obtained from the 1D NPSE is a purely 1D function of $z$ that satisfies the boundary conditions imposed by the 1D Hamiltonian above. In this sense, $\sigma$ functions effectively as an additional variational degree of freedom that is optimised to minimise the ground-state energy, even though its original motivation stems from transverse confinement effects.
	
For these reasons, the 1D NPSE serves here as a direct and appropriate benchmark for assessing the accuracy of the RDM method relative to the standard mean-field approach (the 1D GPE). Both the 1D NPSE and the 2-RDM yield improvements over the 1D GPE (as shown below), but through distinct mechanisms: the former via an extended variational \textit{ansatz} within a 1D framework, and the latter via a beyond-mean-field treatment of particle interactions. This comparison therefore directly tests the capacity of the RDM approach to capture the ground state beyond what is achievable with a mean-field description.
	
	In this work, we compare the ground-state properties derived from the 2-RDM methodology against the 1D GPE and the 1D NPSE in the limit for a large number of bosons. We adopt a system of units for which $\hbar=m=\omega_z=1$, in which case energies are expressed in units of $\hbar\omega_z$ and lengths in units of $a_z=\sqrt{\hbar/(m\omega_z)}$. Moreover, we define $\gamma\coloneqq\omega_{\perp}/\omega_z$ and in these units we note that $g_{\mathsf{3D}}(N-1)\to 4\pi a_{\mathsf{s}}(N-1)$ (where $a_{\mathsf{s}}$ is the \textit{dimensionless} $s$-wave scattering length). Hence, the 1D GPE becomes,
	\begin{equation}
		i\partial_t\varphi	=	\left\{-\frac{1}{2}\triangle_z+\frac{1}{2}z^2+2a_s(N-1)\gamma|\varphi|^2\right\}\varphi,\label{equation:1d_gpe}
	\end{equation}
	and the 1D NPSE becomes
	\begin{align}
		i\partial_t\varphi		&= \left\{-\frac{1}{2}\triangle_z+\frac{1}{2}z^2+\frac{2a_s(N-1)}{\sigma^2}|\varphi|^2 +\frac{1}{2\sigma^2}+\frac{\gamma^2\sigma^2}{2}\right\}\varphi,\label{equation:1d_npse}\\
		0&=1+2a_s(N-1)|\varphi|^2-\gamma^2\sigma^4,\label{equation:1d_npse_alt}
	\end{align}
	where we note that lengths are now measured in terms of $a_z\coloneqq\sqrt{\hbar/(m\omega_z)}$, energy in terms of $\hbar\omega_z$ and time in units of $\omega_z^{-1}$. For convenience, we also make the definitions
	\begin{align}
		\beta\coloneqq 2a_{\mathsf{s}}(N-1)\gamma \qquad \mathsf{and} \qquad g_{\mathsf{1D}}\coloneqq \frac{\beta}{(N-1)}.
	\end{align}
	The equivalence between $\beta/(N-1)$ and $g_{\mathsf{1D}}$ allows us to ensure consistency when comparing results derived from the 2-RDM method and the 1D GPE and its variations. Additionally, we can make a direct comparison between the 1D GPE and 2-RDM results (which strictly model the 1D systems) and the 1D NPSE results (which, as mentioned above, is derived for quasi-1D systems but is aptly applied to more accurately model 1D systems) through the inclusion of $\gamma$ in the interaction strength $\beta$.
	
	\section{Results}
	
		We consider a bosonic system composed of a variable number of particles and, when comparing to the 1D GPE and 1D NPSE, assume $\gamma=10$ in all cases. Moreover, we have used a basis of 20 harmonic oscillator states, and applied the \textsf{D} and \textsf{G} conditions in all cases.
		
		\subsection{Ground state Energies}	
		
\begin{figure}[h]
			\centering
			\includegraphics[scale=1]{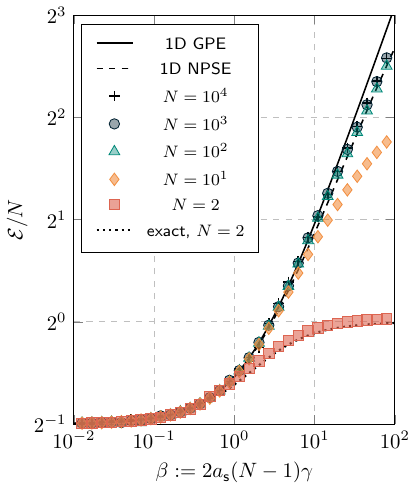}
			\caption{The ground-state energy per boson, as a function of the parameter $\beta$, for the 1D GPE and 1D NPSE, as defined in Eqs.~\eqref{equation:1d_gpe}, \eqref{equation:1d_npse}, and \eqref{equation:1d_npse_alt} along with the ground-state energy derived from the 2-RDM whilst enforcing the \textsf{D} and \textsf{G} conditions, for $N=2,10^1,10^2,10^3$, and $N=10^4$. In each case, $\gamma=10$, and the scattering length can be determined for each $N$ based on the value of $\beta$. The analytic Busch solution \cite{Busch1998} is also provided by the dotted line.}
			\label{figure:ground_state_energy}
		\end{figure}

			In Fig.~\ref{figure:ground_state_energy} we plot the ground state energy $\mathcal{E}/N$ of the system as a function of the interaction strength $\beta$ for a range of boson numbers, $N=2,10,10^2,10^3,$ and $10^4$. In each case of $N$ and $\beta$ the scattering length $a_s$ can be determined, noting that $\gamma=10$. The results show the excellent agreement between the 2-RDM methodology and the analytic Busch solution, demonstrating the expected plateau of the energy in the limit that $\beta$ is large. Moreover, the 2-RDM aligns with the 1D GPE for small interaction strengths, yet agrees with the far more accurate 1D NPSE as the interactions become larger for $N=10^2$ to $10^4$ bosons. The intermediate case of $N=10$ bosons demonstrates the capacity of the 2-RDM method to capture the energy in the intermediate particle number regime.
			
			Fig.~\ref{figure:ground_state_energy} demonstrates the universality of the 2-RDM method in describing systems of particles from $N=2$ to $N=10^4$. Moreover, it allows us to identify the regime in which mean-field approaches become accurate. For example, for $\beta >1$, the mean-field approach is poor at approximating the ground-state energy for a moderate number of bosons ($N\sim 10$) yet, when $N\ge 100$ the mean-field results, particularly when modelled by the 1D NPSE, agree with the results derived from 2-RDM calculations. It is noted further that the 2-RDM results, spanning the few- to many-boson regime are achieved with $20$ basis functions. A larger basis set would result in higher accuracy of the results, yet the applicability of the 2-RDM method is clearly adequately demonstrated in our basis set. This is consistent with results, with similarly sized basis sets, which have have been obtained for a bosonic system with harmonic interactions \cite{Gidofalvi2004}.	
			
\begin{figure}[h]
			\centering
			\includegraphics[scale=1]{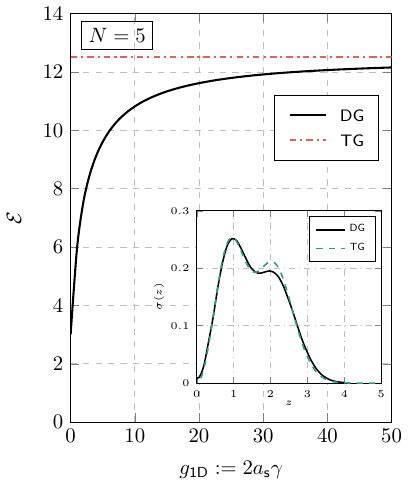}
			\caption{The ground-state energy $\mathcal{E}$ of the $N=5$ boson system, as a function of the parameter $g_{\mathsf{1D}}$, as derived from the 2-RDM method (solid line). We also plot the energy of the TG gas for the $N=5$ system as given in Eq.~\eqref{equation:TG_energy} (dash-dotted line). The inset shows the diagonal pair correlation function $\sigma(z,0)$, as defined in Eq.~\eqref{equation:correlation_function}, as a function of $z$ for the analytic TG wavefunction (dashed line) and that derived from the 2-RDM method (solid line) for $g_{\mathsf{1D}}=50$.}
			\label{figure:ground_state_energy_tonks}
		\end{figure}

			In Fig.~\ref{figure:ground_state_energy_tonks} we also demonstrate the capacity of the 2-RDM methodology to capture the complex attributes of systems with strong interactions. We consider a gas of five bosons and also consider the limit of a TG gas where the harmonic trapping is included in the Hamiltonian, corresponding to our system in the case that $g_{\mathsf{1D}},\beta\to\infty$. In the TG limit, the wavefunction is analytic, being equivalent to the non-interacting fermionic state of a single Slater determinant of the lowest $N$ particle states. This is the celebrated Fermi-Bose mapping theorem of Girardeau \cite{Girardeau1960,Yukalov2005} in which `fermionisation' of the bosons occur. In this limit, the ground-state energy is the energy of a system of spinless fermions,
			\begin{align}
				\mathcal{E}_{\mathsf{TG}}=\frac{N^2}{2}.\label{equation:TG_energy}
			\end{align}
			
		It is noted that the energy provided by the 2-RDM method does not quite capture the analytic energy in the limit that $g_{\mathsf{1D}}\to\infty$. This is due to the limited application of $N$-representability conditions: if $N\le 2$, the \textsf{D} and \textsf{G} conditions are sufficient for a complete description of the ground state (in the basis set), yet this is not the case for $N>5$ where higher-order constraints are formally required. These higher-order constraints are not considered in this work, yet the discussion of them in relation to fermionic systems is covered extensively in the literature \cite{Zhao2004,Knight2022}.
			
		In Fig.~\ref{figure:ground_state_energy_tonks} we plot the ground state energy, $\mathcal{E}$, of the $N=5$ system as a function of the interaction strength $g_{\mathsf{1D}}$, with the dash-dotted red line indicating the energy of the equivalent TG state, i.e., $\mathcal{E}=12.5$. Moreover, we demonstrate the extent to which the 2-RDM captures the `fermionisation' of the boson system at large interaction strengths, which is manifest in the pair correlation function reducing to zero at the origin, $\sigma(z,0)\sim 0$ in the limit that $z\to 0$, where
		\begin{align}
			\sigma(z,0)	&\coloneqq	{}^{(2)}D(z,0;z,0)\\
						&=			\sum_{ijkl}{}^{(2)}D^{ij}_{kl}\varphi_i(z)\varphi_j(0)\varphi_k(z)\varphi_l(0).\label{equation:correlation_function}
		\end{align}		 
		In the inset, $\sigma(z,0)$ is plotted as a function of $z$ for the 2-RDM method at $g_{\mathsf{1D}}=50$ and the TG gas. It is evident that the fermionisation is well-captured, as indicated by the suppression of $\sigma(z,0)$ at $z=0$ in the inset, and the convergence of the ground-state energy via the 2-RDM method to the TG limit. The limitations, in both cases, are a consequence of the choice of basis (which does not accurately capture the cusp in the wavefunction that is apparent due to the Dirac $\delta$-function interaction) and the finite basis rank. This basis set limitation will also naturally result in poorer estimates of the energy in strongly interacting regimes, yet this is expected to be small even in this basis, as demonstrated in Fig.~\ref{figure:ground_state_energy}.
			
		\subsection{Densities}	
		
\begin{figure}[h]
				\centering
				\includegraphics[scale=1]{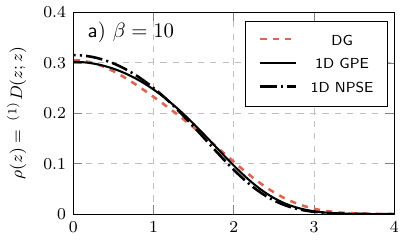}\\
				\includegraphics[scale=1]{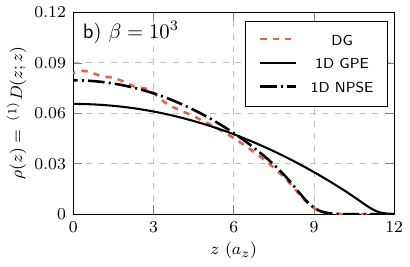}
				\caption{The ground-state density as a function of $z$ for two different values of $\beta$: $\beta=10$ in panel a) and $\beta=10^3$ in panel b). In all cases, $N=10^3$ and $\gamma=10$ and therefore the scattering length in each case can be duly determined. Furthermore, in all cases, we plot the density derived from the 2-RDM methodology (dashed red line), the 1D GPE (solid black line), and 1D NPSE (dash-dotted black line).}
			\label{figure:ground_state_density}
			\end{figure}

			The ground-state energy of our boson system is highly degenerate and we therefore also analyse structural aspects of the ground-state, such as the density, to test the 2-RDM methodology. The density is derivable from the 1-RDM according to
			\begin{align}
				\rho(z)	&\coloneqq	{}^{(1)}D(z,z)\\
						&=\sum_{ij}{}^{(1)}D^i_j\varphi_i^*(z)\varphi_j(z).
			\end{align}
			In Fig.~\ref{figure:ground_state_density} we plot the ground-state density of the $N=10^3$ system for two different interaction strengths $\beta=10$ and $10^3$, where $\gamma=10$ in all cases. The density as derived from the 2-RDM methodology is given by the dashed red line, the density from the 1D GPE by the solid black line, and the density from the 1D NPSE (dash-dotted black line).
			
			The interaction strengths in Fig.~\ref{figure:ground_state_density} cover two separate regimes: relatively weakly interacting at $\beta=10$ and strongly interacting at $\beta=10^3$. In the former, the density obtained using the 2-RDM methodology, the 1D GPE, and the 1D NPSE agree, indicating that the 2-RDM method accurately reflects in the mean-field result for appropriate interaction strengths. However, for the latter case, where the interactions begin to dominate the ground-state and the 1D GPE fails to capture the density, the 2-RDM density agrees well with the density derived from the 1D NPSE. These results collectively demonstrate the capacity of the 2-RDM methodology to capture structural properties of the ground-state for large particle numbers (i.e., $N=10^3$) and up to large interaction strengths $\beta=10^3$.
			
			We note that, due to the use of the limited basis of rank $20$ for the 2-RDM methodology, very small ripples appear within the density, particularly for panel b) in Fig.~\ref{figure:ground_state_density}; it is expected that in the limit of a large basis rank these ripples would disappear. Notably, these are not structural ripples that reflect complex phenomenology within the ground-state, as, for example, indicated in the correlation function in the TG gas in Fig.~\ref{figure:ground_state_energy_tonks}. This is because $N=10^3$ is sufficiently large to minimise these effects within the density profile.
		
		\subsection{Correlations}
		
			The correlation function was plotted in one instance for the $N=5$ system in Fig.~\ref{figure:ground_state_energy_tonks} and here we analyse the correlation function in more detail. The correlation function provides us with details pertaining to the pair distribution of bosons, giving a numerical calculation of the likelihood of bosons being in proximity (it becomes a direct probability distribution function for the liklihood of a particle at $z$ and another at $z=0$ when appropriately normalised). In strongly interacting systems, such as the TG gas, the pair correlation function is naturally expected to be suppressed when the coordinates of bosons coincide. 
			
\begin{figure}[h]
				\centering
				\includegraphics[scale=1]{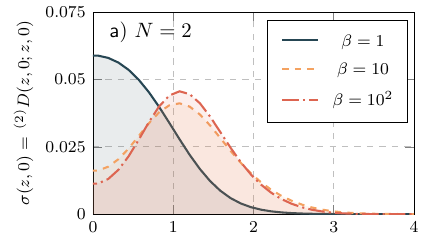}\\
				\includegraphics[scale=1]{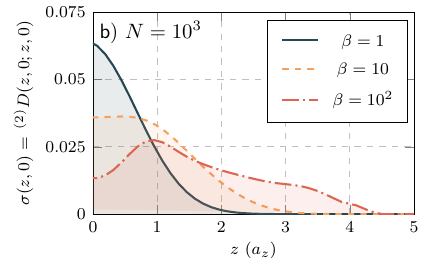}\\
				\caption{The ground-state correlation, $\sigma(z,0)$, as derived from the 2-RDM, as a function of $z$ for $N=2$ (panel a)) and $N=10^3$ (panel b)) and, in each case, for three different values of $\beta$: $\beta=1$ given by solid blue line, $\beta=10^2$ given by dashed orange line, and $\beta=10^3$ given by dash-dotted red line. In all cases, $\gamma=10$ and therefore the scattering length in each case can be duly determined.}
			\label{figure:ground_state_correlation}
			\end{figure}

			In Fig.~\ref{figure:ground_state_correlation} we plot the pair correlation function $\sigma(z,0)$, which provides the likelihood of a boson being located at $z$ given another is located at the origin. This function is plotted for both $N=2$ and $N=10^3$ for a range of different interaction strengths $\beta=1,10,100$. In both cases, we identify a lack of suppression for small interaction strengths at the origin, and a large suppression as the interaction strength increases. Moreover, this suppression is similar, both for the $N=2$ and $N=10^3$ cases; we note, however, that the suppression at $\beta=10$ for the $N=10^3$ case is less than for the $N=2$ case. This pronounced difference in the suppression of $\sigma(z,0)$ at $\beta=10$ is because this value lies in an intermediate interaction regime. For small particle numbers ($N = 2$), the effective interaction strength is much larger at fixed $\beta$, so the system is already deep in the strongly interacting regime at $\beta = 10$ and shows strong suppression of the pair correlation at the origin. For $N = 10^3$, however, the effective coupling per particle is reduced by roughly three orders of magnitude; at $\beta = 10$ the system is therefore only moderately interacting and the suppression remains relatively weak. At the lowest interaction strength ($\beta = 1$) both systems are weakly interacting and exhibit little suppression, while at the highest strength ($\beta = 10^2$) both are firmly in the strongly interacting regime with strong suppression near $z = 0$. Consequently, the difference between the few- and many-body curves is maximised in the intermediate regime around $\beta = 10$.

\begin{figure}[h]
				\centering
				\includegraphics[scale=1]{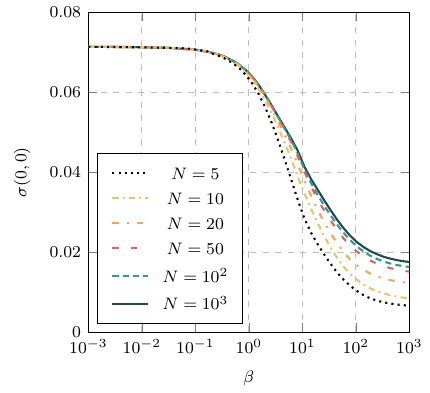}
				\caption{The value of the correlation function $\sigma(z,0)$ at the origin, i.e., $\sigma(0,0)=\lim_{z\to 0}\sigma(z,0)$ as a function of $\beta$ for $N=5$ to $N=10^3$ as indicated. In all cases, $\gamma=10$.}
			\label{figure:ground_state_correlation2}
			\end{figure}

		The $N=2$ and $N=10^3$ cases constitute two extreme regimes, the few-body $(N=2)$ case with an analytic solution, and the $N=10^3$ case where mean-field descriptions are typically used. A key test of the 2-RDM methodology is to study the crossover regime between these two cases, and identify if continuity is manifest in observable quantities. In Fig.~\ref{figure:ground_state_correlation2} we plot the value of $\lim_{z\to0}\sigma(z,0)\coloneqq \sigma(0,0)$ (normalised to unity) as a function of the interaction strength $\beta$ for a range of boson numbers $N=5,10,20,50,10^2,$ and $10^3$. The plot demonstrates the continuous evolution of $\sigma(0,0)$ as $N$ increases, with no irregularities occurring within the crossover regime from few to many bosons. The consistency across $N$ and $\beta$ indicates that the 2-RDM methodology successfully captures the key physical behaviours inherent to both few- and many-body systems.
		
			Of particular importance, the observed behaviour in the crossover regime ($N\sim 100$) indicates that the 2-RDM methodology is robust in the regime where mean-field approaches are potentially inaccurate, but typical methods used for few-body systems are intractable. As such, it can be argued that the 2-RDM methodology is universal, capable of capturing the properties of the ground-state across regimes that typically require a suite of different methods.
			
	\section{Concluding Remarks}
	
		In this work, we have demonstrated the efficacy of the 2-RDM methodology in describing bosonic systems comprising few to many particles. The 2-RDM method has seen widespread success in the study of electronic structure in atomic and molecular systems yet has had limited utilisation for bosons. Moreover, the 2-RDM method had only been applied to systems consisting of short-range `contact' interactions in our recent work \cite{Knight2022}. As such, due to the lack of size-extensiveness of the 2-RDM, and the absence of the Pauli exclusion principle which places natural constraints on the size of basis used, the 2-RDM method has been applied to large systems comprising $N=10^3$ to $10^4$ bosons across a range of (repulsive) interaction strength regimes.
	
		Specifically, we demonstrate that the ground-state energy calculated via a variational procedure utilising the 2-RDM accurately reproduces the analytic results in the $N=2$ case, and aligns extremely well with highly accurate mean-field methods derived by Salasnich \cite{Salasnich2002} for large particle numbers and strong interaction strengths. The 2-RDM methodology therefore successfully captures the ground-state energy across a range of extreme regimes, from small to large $N$ and $\beta$.
		
		Additionally, the 2-RDM methodology accurately reproduces the density as compared to the highly accurate mean-field descriptions for $N=10^3$; this accuracy is retained for large interaction strengths where the `standard' 1D GPE fails dramatically. Moreover, the correlation function for $N=2$ and $N=10^3$ is obtained, with the suppression of the pair-correlation at the origin for large interaction strengths demonstrated. These results collectively demonstrate the capacity of the 2-RDM method to capture the structural properties of the ground state for few-to many-boson systems.
		
		A crucial test, however, is the capacity of the 2-RDM method to accurately obtain ground-state properties in the crossover regime from few-to many-boson systems. We demonstrate capacity by plotting the pair correlation function at the origin for a range of values of $N$, from $N=5$ to $N=10^3$. We identify a smooth transition of this quantity from the $N=5$ to the $N=10^3$ case, without any irregularities. As such, the crossover regime is well-captured.
		
		The 2-RDM method is well-suited to the description of bosonic systems regardless of the number of bosons. Notably, only the \textsf{D} and \textsf{G} conditions have been applied yet accurate results have been obtained in the $N=2$ case, where a comparison to the analytic solution is given, and the $N=10^3$ and $N=10^4$ case, where the RDM method demonstrates improvement on the mean-field result. The structural properties in the intermediate regime, where $N=5$, have also shown good agreement with the analytic result in the strongly interaction regime. As mentioned, this applicability of the RDM method to bosonic systems was also demonstrated by Gidofalvi \textit{et al} \cite{Gidofalvi2004} for a different one-dimensional system. Further tests of this method can be conducted on relatively simple systems. For example, an immediate system to study is the same system studied in this work yet with attractive interactions, where it is expected that the RDM methodology will yield similarly accurate results. It is noted, however, that the efficacy of the application of the RDM in this context will heavily depend upon the basis, with the quantum harmonic oscillator basis potentially failing to capture the complicated structure of the ground state. For example, the presence of solitons and droplets in 1D bosonic systems, as well as comparisons in the 3D case generally against the 3D GPE are interesting targets. Moreover, this method could be utilised to test the accuracy of the mean-field approaches for many-boson systems wherein there is disagreement with experiment. For example, corrections, nominally arising from quantum fluctuations, to the 3D GPE are typically made to account for the phenomenology of droplets in dipolar BECs \cite{FerrierBarbut2016}. The 2-RDM method could therefore be utilised to more accurately describe this correction.	

	\section{Acknowledgements}

		M. J. K. was supported by an Australian Government Research Training Program Scholarship and by the University of Melbourne and numerical results were obtained using the Spartan high-performance computing system at the University of Melbourne \cite{Meade2017}.

\bibliography{reduced_density_matrix_approach_to_one_dimensional_bose_einstein_condensates.bib}
%
%
%
%
\end{document}